\documentclass[letterpaper, 10 pt, conference]{ieeeconf}  % Comment this line out if you need a4paper 

\IEEEoverridecommandlockouts                              % This command is only needed if 
\usepackage[dvipdfmx]{graphicx}
\usepackage{epsfig} % for postscript graphics files
\usepackage{mathptmx} % assumes new font selection scheme installed
\usepackage{times} % assumes new font selection scheme installed
\usepackage{amsmath} % assumes amsmath package installed
\usepackage{amssymb}  % assumes amsmath package installed

\usepackage{algorithm}
\usepackage{algorithmicx}
\usepackage[noend]{algpseudocode}
\usepackage{color}
\usepackage{xcolor}
\usepackage{url}
\usepackage{caption}

\DeclareMathAlphabet{\mathcal}{OMS}{cmsy}{m}{n}

\title{\LARGE \bf
MARPF: Multi-Agent and Multi-Rack Path Finding
}

\author{Hiroya Makino$^{1,*}$, Yoshihiro Ohama$^{1}$, and Seigo Ito$^{1}$% <-this % stops a space
\thanks{$^{1}$ H. Makino, Y. Ohama, and S. Ito are with Toyota Central R\&D Labs., Inc., 41-1, Yokomichi, Nagakute, Aichi, Japan.}% \\
        \thanks{$^{*}$ Corresponding author. {\tt\small hirom@mosk.tytlabs.co.jp}}%
        }

\begin{document}

\maketitle
\thispagestyle{empty}
\pagestyle{empty}

\renewcommand{\thefootnote}{\fnsymbol{footnote}}
\footnote[0]{© 2024 IEEE.  Personal use of this material is permitted.  Permission from IEEE must be obtained for all other uses, in any current or future media, including reprinting/republishing this material for advertising or promotional purposes, creating new collective works, for resale or redistribution to servers or lists, or reuse of any copyrighted component of this work in other works.}
\renewcommand{\thefootnote}{\arabic{footnote}}

%%%%%%%%%%%%%%%%%%%%%%%%%%%%%%%%%%%%%%%%%%%%%%%%%%%%%%%%%%%%%%%%%%%%%%%%%%%%%%%%
\begin{abstract}
In environments where many automated guided vehicles (AGVs) operate, planning efficient, collision-free paths is essential. 
Related research has mainly focused on environments with pre-defined passages, resulting in space inefficiency. 
We attempt to relax this assumption.
In this study, we define multi-agent and multi-rack path finding (MARPF) as the problem of planning paths for AGVs to convey target racks to their designated locations in environments without passages.
In such environments, an AGV without a rack can pass under racks, whereas one with a rack cannot pass under racks to avoid collisions.
MARPF entails conveying the target racks without collisions, while the obstacle racks are relocated to prevent any interference with the target racks.
We formulated MARPF as an integer linear programming problem in a network flow.
To distinguish situations in which an AGV is or is not loading a rack, the proposed method introduces two virtual layers into the network.
We optimized the AGVs' movements to move obstacle racks and convey the target racks.
The formulation and applicability of the algorithm were validated through numerical experiments.
The results indicated that the proposed algorithm addressed issues in environments with dense racks. 
\end{abstract}

%%%%%%%%%%%%%%%%%%%%%%%%%%%%%%%%%%%%%%%%%%%%%%%%%%%%%%%%%%%%%%%%%%%%%%%%%%%%%%%%

\section{Introduction}
\label{introduction}
Over the past few decades, introducing automated guided vehicles (AGVs) in warehouses and factories has accelerated efficiency.
Numerous efforts have been devoted to developing multi-agent path finding (MAPF) for efficient transportation via AGVs \cite{stern2019}.
This problem has been applied in many fields, such as automatic warehouses \cite{wurman2008, honig2019}, airport taxiway control \cite{li2019}, and automated parking \cite{okoso2022}.

\begin{figure}[!t]
    \centering
    \includegraphics[scale=0.95]{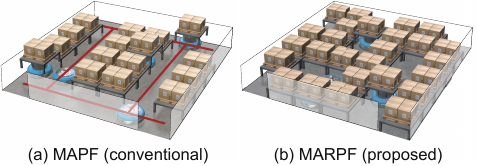}
    \caption{Environments of MAPF and MARPF. MAPF needs passages, whereas MARPF does not.}
    \label{figure1}
\end{figure}

For transportation in warehouses and factories, AGVs often navigate beneath rack-type carts and convey entire racks with their contents (hereinafter referred to as rack) to a designated location.
Prior research on MAPF has mainly considered navigating these racks through areas with passages (Fig. \ref{figure1}(a)). 
However, this layout leads to the inefficient use of space. 
In settings where sufficient space cannot be provided, it is crucial to utilize the available area more effectively. 
Traditional MAPF algorithms struggle to optimize navigation racks efficiently in these dense environments (Fig. \ref{figure1}(b)). 
The difficulty lies in optimally relocating the obstacle racks.

This study defines multi-agent and multi-rack path finding (MARPF), which focuses on planning paths for AGVs to convey target racks to their designated locations in a grid-like environment without passages. 
The racks cannot move and, thus, should be conveyed by AGVs.
AGVs without racks can pass under the racks; however, to avoid collisions, those with racks cannot.
MARPF involves conveying the target racks to their designated destinations, while the obstacle racks are situated freely.
In an environment in which racks are densely located, the obstacle racks can be moved to avoid interference with the target racks.

\begin{figure}[!t]
  \centering
  \includegraphics[scale=0.95]{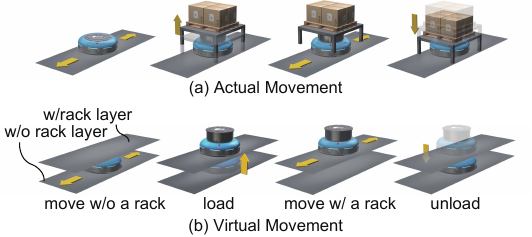}
  \caption{Two layers to express AGVs’ movements. The movements between these layers correspond to whether they are conveying a rack or not. When an AGV loads a rack, it virtually transitions from the ``w/o rack'' layer to the ``w/rack'' layer; unloading a rack reverses this transition.}
  \label{two_layers}
\end{figure}

We formulated MARPF as an integer linear programming (ILP) problem in a network flow, which includes the AGV and rack networks.
In the AGV network, the proposed method distinguishes whether an AGV is conveying a rack using two virtual layers (Fig. \ref{two_layers}).
The rack network represents the movements of the racks, which are separated from the AGVs.
By synchronizing the AGV network with the rack network, the proposed method enables moving obstacle racks and conveying target racks while avoiding collisions.
We aimed to solve the problem with various movement constraints and minimize the makespan (i.e., the latest completion time).

\subsection{Contribution}
Original multi-agent pickup and delivery (MAPD) focuses only on conveying target racks. Other studies have considered switching the positions of racks \cite{li2023, bachor2023}.
The difference between MARPF and rearrangement-considering MAPD is as follows:
\begin{itemize}
  \item \textbf{MARPF}: Only the target racks are assigned goals; the obstacle racks are situated freely.
  \item \textbf{Rearrangement-considering MAPD}: All racks are assigned goals.
\end{itemize}

\begin{figure}[!t]
  \vspace*{1.35mm}
  \centering
  \includegraphics[scale=0.95]{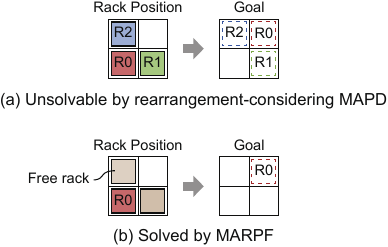}
  \caption{Unsolvable example by MAPD.}
  \label{dd_mapd_example}
\end{figure}

It could be posited that rearrangement-considering MAPD can potentially address MARPF by assigning the same goal positions as the initial positions for obstacle racks. 
However, we estimate  two typical situations where this assumption may not apply: (1) Environments with highly dense racks, where some situations cannot be solved by rearrangement-considering MAPD (Fig. \ref{dd_mapd_example}). Sufficient conditions for solvability are discussed in a sliding tile puzzle \cite{johnson1879}; and (2) Environments with dense racks and a limited number of AGVs, where relocating racks to their initial positions would require a significant number of time steps.
The optimization problem of where to relocate the obstacle racks is challenging, and MARPF constitutes a new problem.

The main contributions of this study are as follows:
\begin{itemize}
  \item We define a new problem (MARPF) of planning paths for AGVs to convey target racks to their designated locations in dense environments.
  \item We propose a method for solving MARPF, formulated as an ILP problem in a network flow.
  \item Solving the full ILP problem would be computationally intractable; therefore, we propose a hybrid approach combined with cooperative A* (CA*) to allow feasible computation times.
\end{itemize}

\subsection{Related Work}
\subsubsection{MAPF}
The MAPF problem relates to finding the optimal paths for multiple agents without collisions, and numerous methods have been proposed \cite{stern2019}.
As complete and optimal solvers, there are conflict-based search (CBS) \cite{sharon2015}, improved CBS \cite{boyarski2015}, and enhanced CBS \cite{barer2014}.
Prioritized planning, such as CA* \cite{silver2005} and multi-label A* \cite{grenouilleau2021}, has a short runtime but is suboptimal.
MAPD \cite{ma2017} is a lifelong variant of MAPF.
Specifically, in MAPD, each agent is constantly assigned new tasks with new goal locations, whereas in MAPF, each agent has only one task.

\subsubsection{Rearrangement-considering MAPD}
In Double-Deck MAPD (DD-MAPD) \cite{li2023}, agents are tasked to move racks to their assigned delivery locations, thereby changing the overall arrangement of the racks.
The algorithm for DD-MAPD solves a DD-MAPD instance with $M$ agents and $N$ racks by first decomposing it into an $N$-agent MAPF instance, and subsequently into a $M$-agent MAPD instance with task dependencies.
In Multi-Agent Transportation (MAT) \cite{bachor2023}, all racks are assigned delivery locations without fixed aisles.
An algorithm is provided for solving MAT by reducing it to a series of satisfiability problems.
In DD-MAPD and MAT, all racks must be assigned goals.
However, they do not consider where and how obstacle racks are relocated, which is necessary for MARPF.

\section{Problem Definition}
In this section, we define the MARPF problem.
As noted in Section \ref{introduction}, MARPF aims to plan paths for AGVs to convey target racks in an environment without passages.
Table \ref{notation} lists the notation used in the following sections.

An MARPF instance comprises $M$ AGVs, $N$ racks, and an undirected grid graph, $G=(V,E)$.
\begin{align}
  &V=\{v_{x,y} \mid x \in \{0, \ldots, size_x-1\}, y \in \{0, \ldots, size_y-1\}\}, \nonumber\\
  &E=\{(v_{x,y}, v_{x^\prime,y^\prime}) \mid v_{x,y}, v_{x^\prime,y^\prime} \in V, |x-x^\prime|+|y-y^\prime|=1\}, \nonumber
\end{align}
where $v_{x,y} \in V$ is the vertex with a column index $x$ and row index $y$ in the grid.
Diagonal edges are not considered to avoid collisions between AGVs and the legs of racks. 
$loc^t(a_i), loc^t(r_i) \in V$ denote the vertices of AGV $a_i$ and rack $r_i$ at timestep $t$, respectively.
$cnv^t(a_i) \in \{0,1\}$ refers to the rack-conveying state of AGV $a_i$ at timestep $t$, where $1$ indicates that the AGV is conveying a rack, and $0$ indicates otherwise.

Time is assumed to be discretize.
At each time step $t$, an AGV executes one of the following actions:
\begin{enumerate}
    \item Remain at the current location. \\$loc^t(a_i)=loc^{t+1}(a_i)$.
    \item Move to the next location. \\$\left(loc^t(a_i), loc^{t+1}(a_i) \right) \in E$.
    \item Load a rack. \\$cnv^t(a_i) = 0$, $cnv^{t+1}(a_i) = 1$, $loc^t(a_i)=loc^{t+1}(a_i)$.
    \item Unload a rack. \\$cnv^t(a_i) = 1$, $cnv^{t+1}(a_i) = 0$, $loc^t(a_i)=loc^{t+1}(a_i)$.
\end{enumerate}
A rack also executes one of the following actions: remain $loc^t(r_i)=loc^{t+1}(r_i)$ or move $\left(loc^t(r_i), loc^{t+1}(r_i) \right) \in E$; however, its movement depends on AGVs because the rack does not move by itself.

\begin{table}[!t]
  \vspace*{1.35mm}
  \small
  \centering
  \captionsetup{justification=centering,singlelinecheck=false}
  \caption{Notations.}
  \begin{tabular}{ll}
    \hline
    Symbols                        & Description                                                     \\
    \hline
    \multicolumn{2}{l}{Section II and after}                                     \\
    $a_i$                          & AGV $i$                                                         \\
    $r_i$                          & Rack $i$                                                         \\
    $x\in \{0, \ldots, size_x-1\}$  & Column index ($size_x$ is the grid width)                 \\           
    $y\in \{0, \ldots, size_y-1\}$  & Row index ($size_y$ is the grid height)                                                 \\
    $v_{x,y}$                      & Vertex  \\                                              
    $V$                            & Set of vertices                                                     \\
    $E$                            & Set of edges                                                     \\
    $G=(V,E)$                      & Connected undirected graph \\
    $t$                  & Timestep \\
    $loc^t(a_i)$           & Location vertex of $a_i$ at timestep $t$                                      \\
    $cnv^t(a_i)$                  & Rack-conveying state of $a_i$\\
                     & at timestep $t$\\\hline
    \multicolumn{2}{l}{Section III and after}                                     \\
    $G^{+}_{ag}=(V^{+}_{ag}, E^{+}_{ag})$      & Time-expanded network representing \\
                                   & all AGVs' movements                                          \\
    $G^{+}_{ar}=(V^{+}_{ar}, E^{+}_{ar})$      & Time-expanded network representing \\
                                   & all racks' movements                                           \\
    $G^{+}_{tr}=(V^{+}_{tr}, E^{+}_{tr})$ & Time-expanded network representing\\
                                   & the target rack's movements                                     \\
    $\mathcal{T}$                  & Set of timesteps in time-expanded \\
                                   & networks\\
    $c(\cdot)$                     & Cost function of flow                                                       \\
    $l$                    & Layer (w/rack or w/o rack) \\
    $S_{ag}, S_{ar}, S_{tr} \subseteq V$       & Start location vertices of all AGVs, \\
                                   & all racks, and the target racks, \\
                                   & respectively                                  \\
    $T_{tr} \subseteq V$                       & Goal location vertex of the target rack                                                 \\
    $\alpha^t_{vi, vo, li, lo}, \beta^t_{vi, vo}, \gamma^{t}_{vi, vo}$    & Flow on $E^{+}_{ag}, E^{+}_{ar}, E^{+}_{tr}$, respectively \\ \hline
  \end{tabular}
  \label{notation}
\end{table}

\begin{figure}[!t]
  \vspace*{1.35mm}
  \centering
  \includegraphics[scale=0.95]{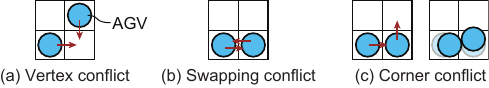}
  \caption{Type of conflicts.}
  \label{conflict}
\end{figure}

There should be no collisions between agents or racks.
We define three types of collision for agents (Fig. \ref{conflict}): 
The first is the vertex conflict \cite{stern2019}, where two agents cannot be in the same location at the same timestep. 
The second is the swapping conflict \cite{stern2019}, where two agents cannot move along the same edge in opposite directions at the same timestep.
The third is the corner conflict. 
When an agent moves to the other agent's position and their adjacent agent's movements are in a vertical relationship, their corners collide, as shown in Fig. \ref{conflict}(c).

Formally, for all agents $a_i, a_j \, (i \neq j)$ and all timesteps $t$, the following equations must hold to avoid collisions:
\begin{align}
  &loc^t(a_i) \neq loc^t(a_j), \label{def:vertexconflict}\\
  &loc^t(a_i) = loc^{t+1}(a_j) \nonumber\\
  &\Rightarrow \left( loc^{t+1}(a_i)_x - loc^t(a_i)_x \right) = \left( loc^{t+1}(a_j)_x - loc^t(a_j)_x \right) \nonumber\\
  & \quad \wedge \left( loc^{t+1}(a_i)_y - loc^t(a_i)_y \right) = \left( loc^{t+1}(a_j)_y - loc^t(a_j)_y \right), \label{def:cornerconflict}
\end{align}
where $loc^t(a_i)_x$ and $loc^t(a_i)_y$ denote the column and row indices of $loc^t(a_i)$, respectively.
\eqref{def:vertexconflict} represents the vertex conflict.
\eqref{def:cornerconflict} represents both the swapping and corner conflicts.
These three types of collision are also applied to racks.

The MARPF problem involves computing collision-free paths for AGVs and minimizing the timesteps required to convey the target racks to their designated locations.
Non-target racks act as obstacles and are positioned freely.

\begin{figure}[!t]
    \centering
    \includegraphics[scale=0.95]{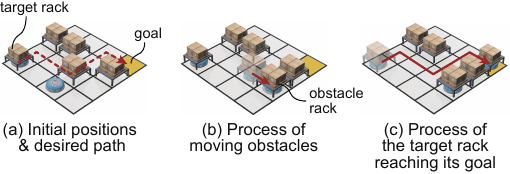}
    \caption{Problem example.}
    \label{problem_example}
\end{figure}

Fig. \ref{problem_example} exemplifies moving a rack to avoid interference with the movement of the target rack.
The path the AGV must follow to convey the target rack to its designated location is blocked by an obstacle rack (Fig. \ref{problem_example}(a)).
First, the obstacle rack is removed (Fig. \ref{problem_example}(b)), and the target rack is then conveyed to the designated location (Fig. \ref{problem_example}(c)).

\section{Proposed Method}
In this section, we formulate MARPF as a minimum-cost flow problem in a network flow, similar to the approaches for MAPF \cite{okoso2022, yu2013}.
Two types of synchronized networks are established, one for AGVs and the other for racks.
The network for AGVs distinguishes whether or not an AGV is conveying a rack using two virtual layers to represent conveying a rack.

\subsection{Definition of Networks}
When AGVs are not conveying a rack, their obstacles are the other AGVs; however, when they are conveying a rack, other racks also become obstacles. 
Therefore, whether an AGV is conveying a rack onto a time-expanded network must be determined.
The AGVs' movements are classified into the following four actions in terms of the rack-conveying state (Fig. \ref{two_layers}(a)):
\begin{enumerate}
  \item Move (or remain at the current location) without a rack.
  \item Load a rack.
  \item Move (or remain at the current location) with a rack.
  \item Unload a rack.
\end{enumerate}

Our proposed method divides a moving plane into two virtual layers to represent two situations: conveying or not conveying a rack (Fig. \ref{two_layers}(b)).

The rack-loading action corresponds to movement from the w/o to the w/rack layers, and vice versa.

\begin{figure*}[!t]
  \vspace*{1.35mm}
  \centering
  \includegraphics[scale=0.95,pagebox=cropbox,clip]{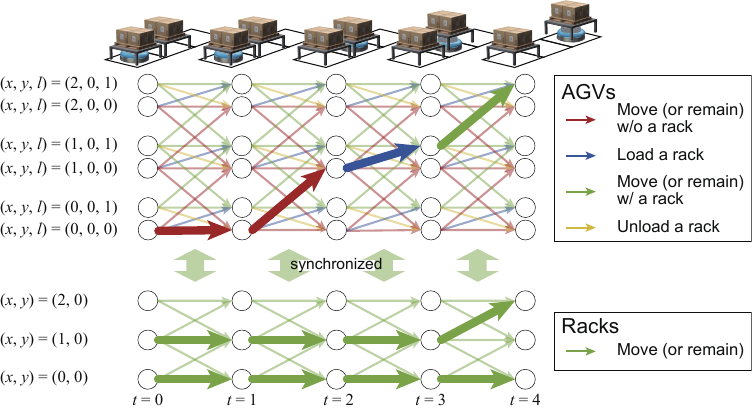}
  \caption{Time-expanded networks for AGVs (upper) and racks (lower). Loading or unloading a rack (blue and yellow arrows, respectively) only changes the loading status $l$, whereas moving (red and green arrows) changes the position $x$ or $y$ but not $l$. The two networks are synchronized, and the movement of racks depends on the movement of AGVs. The movement of the AGV and racks at the top of the figure corresponds to the thick arrows.}
  \label{agv_and_rack_network}
\end{figure*}

Then, all vertices are expanded along the time-axis, and under specific constraints, directed edges are added.
The edges indicate AGV motion in the grid.
The motions are distinguished by the indices $(x, y)$ of the source and sink vertices and the virtual layers, where the motion is with ($l=1$) and without ($l=0$) a rack.
Fig. \ref{agv_and_rack_network} (upper) shows an example of a time-expanded network representing all AGVs' movements.
For simplicity, Figs. \ref{two_layers} and \ref{agv_and_rack_network} show the moving plane in one dimension; however, it is two-dimensional.

For the racks’ movements, the networks should be separated from the AGVs.
Distinguishing the layers in a time-expanded network is not necessary (Fig. \ref{agv_and_rack_network} (lower)).
The flows expressing the movements of racks are synchronized with the corresponding flows of AGVs.

We define the time-expanded network representing all AGVs, all racks, and the target rack's movements as directed graphs $G^{+}_{ag}=(V^{+}_{ag}, E^{+}_{ag})$, $G^{+}_{ar}=(V^{+}_{ar}, E^{+}_{ar})$, and $G^{+}_{tr}=(V^{+}_{tr}, E^{+}_{tr})$, respectively.
$G^{+}_{tr}$ has the same structure as $G^{+}_{ar}$; however, the target rack is distinguished from the other racks.
The goal location of the target rack is fixed on $G^{+}_{tr}$. 
\begin{align}
  &
  \begin{array}{@{}l@{}l@{}}
    V^{+}_{ag}=\{v^t_{x,y,l} \mid \: &x \in \{0, \ldots, size_x-1\}, y \in \{0, \ldots, size_y-1\}, \nonumber\\
    & l\in \{0,1\}, t \in \{0, \ldots, size_t\}\},\nonumber
  \end{array}\\
  &
  \begin{array}{@{}l@{}l@{}}
    E^{+}_{ag}=\{\langle v^t_{x,y,l}, v^{t^\prime}_{x^\prime,y^\prime,l^\prime}\rangle \mid \: &v^t_{x,y,l}, v^{t^\prime}_{x^\prime,y^\prime,l^\prime} \in V^{+}_{ag}, \nonumber\\
    & |x-x^\prime|+|y-y^\prime|\leq 1, t^\prime-t=1\},\nonumber
  \end{array}\\
  &
  \begin{array}{@{}l@{}l@{}}
    V^{+}_{ar}, V^{+}_{tr}=\{v^t_{x,y} \mid \: &x \in \{0, \ldots, size_x-1\}, \nonumber\\
    & y \in \{0, \ldots, size_y-1\}, t \in \{0, \ldots, size_t\}\},\nonumber
  \end{array}\\
  &
  \begin{array}{@{}l@{}l@{}}
    E^{+}_{ar}, E^{+}_{tr}=\{\langle v^t_{x,y}, v^{t^\prime}_{x^\prime,y^\prime}\rangle \mid \: &v^t_{x,y}, v^{t^\prime}_{x^\prime,y^\prime} \in V^{+}_{ar}, \nonumber\\
    & |x-x^\prime|+|y-y^\prime|\leq 1, t^\prime-t=1\},\nonumber
  \end{array}
\end{align}
where $size_t$ denotes the maximum timesteps in the time-expanded network.
The value of $size_t$ is set large enough for the target agent to reach the goal.
Dependencies exist between $G^{+}_{ar}$ and $G^{+}_{ag}$ and between $G^{+}_{tr}$ and $G^{+}_{ar}$, and their flows are synchronized.

\subsection{Definition of Variables}
The vertices on $V^{+}_{ag}$ are distinguished by $t, v$, and $l$.
Each edge consists of the source and sink vertices.
$\alpha^t_{vi, vo, li, lo}$ represents the flow on the edge $\langle v^t_{vi_x,vi_y,li}, v^{t+1}_{vo_x,vo_y,lo}\rangle \in E^{+}_{ag}$.
Here, $vi, vo \in V, li, lo \in \{0,1\}$ respectively denote the source vertex, sink vertex, source layer, and sink layer, and $t \in \mathcal{T}=\{0, \ldots, size_t-1\}$ represents the source vertex's timestep.
$G^{+}_{ar}, G^{+}_{tr}$ do not contain multiple layers; therefore, the flows of racks are expressed more simply as $\beta^t_{vi, vo}, \gamma^t_{vi, vo}$, respectively.
We define the variables and the cost function $c(\cdot)$ as follows:
\begin{align}
    &\alpha^t_{vi, vo, li, lo}, \beta^t_{vi, vo}, \gamma^t_{vi, vo} \in \{0, 1\}, \\
    &c(vi, vo, li, lo, t) = 
    \left\{
    \begin{array}{ll}
    0 &\text{if } \, vi=vo, li=lo \\
    t+1 &\text{otherwise}
    \end{array}
    \right. .
\end{align}

We define the initial location vertices of all AGVs, racks, and the target rack, respectively, as $S_{ag}, S_{ar}$, and $S_{tr}$, where $S_{tr} \subseteq S_{ar}$.
In the network, we express the initial locations by fixing the flows at $t=0$ (respectively, \eqref{1_alpha_start}, \eqref{1_beta_start}, and \eqref{1_gamma_start}).
The goal location vertex of the target rack, $T_{tr}$, is defined by \eqref{1_gamma_goal}.
\begin{align}
&\left\{
\begin{array}{ll}
\alpha^0_{vi, vo, li, lo} = 1 &\text{if } \, vi=vo \in S_{ag}, li=lo=0 \\
\alpha^0_{vi, vo, li, lo} = 0 &\text{otherwise}
\end{array}
\right. \label{1_alpha_start} \\
&\left\{
\begin{array}{ll}
\beta^0_{vi, vo} = 1 &\text{if } \, vi=vo \in S_{ar} \\
\beta^0_{vi, vo} = 0 &\text{otherwise}
\end{array}
\right. \label{1_beta_start} \\
&\left\{
\begin{array}{ll}
\gamma^0_{vi, vo} = 1 &\text{if } \, vi=vo \in S_{tr}\\
\gamma^0_{vi, vo} = 0 &\text{otherwise}
\end{array}
\right. \label{1_gamma_start} \\
&\left\{
\begin{array}{ll}
\gamma^{size_t-1}_{vi, vo} = 1 &\text{if } \, vi=vo \in T_{tr} \\
\gamma^{size_t-1}_{vi, vo} = 0 &\text{otherwise}
\end{array}
\right. \label{1_gamma_goal}
\end{align}

\subsection{Minimum Cost Flow Problem}
\label{mcfp}
The aim is to convey the target rack from the starting position to the goal position as quickly as possible.
The set of timesteps $\mathcal{T}\setminus \{ size_t-1\}$ refers to $\mathcal{T}^{-}$.
\begin{flalign}
    &\text{min} \, \sum c(vi, vo, li, lo, t)\alpha^t_{vi, vo, li, lo} &&  \label{eq:obj_func}\\
    &\text{s.t.} && \nonumber \\
    &\sum_{vi, li} \alpha^t_{vi, v, li, l} = \sum_{vo, lo} \alpha^{t+1}_{v, vo, l, lo}  &&  \forall v\in V, l \in \{0, 1\}, t\in \mathcal{T}^{-}\label{eq:i_flow_o_flow_alpha}\\
    &\sum_{vi} \beta^t_{vi, v} = \sum_{vo} \beta^{t+1}_{v, vo} &&  \forall v\in V, t\in \mathcal{T}^{-} \label{eq:i_flow_o_flow_beta}\\
    &\sum_{vi} \gamma^t_{vi, v} = \sum_{vo} \gamma^{t+1}_{v, vo} &&  \forall v\in V, t\in \mathcal{T}^{-} \label{eq:i_flow_o_flow_gamma}\\
    &\beta^t_{vi, vo} = \alpha^t_{vi, vo, 1, 1} &&  \forall (vi, vo) \in E, t \in \mathcal{T} \label{eq:beta_depends_alpha}\\
    &\gamma^t_{vi, vo} \leq \beta^t_{vi, vo} &&  \forall vi, vo \in V, t \in \mathcal{T} \label{eq:gamma_depends_beta}\\
    &\alpha^t_{v, v, 0, 1} \leq \sum_{vi} \beta^t_{vi, v} &&  \forall v \in V, t \in \mathcal{T} \label{eq:alpha_load}\\
    &\sum_{vi, li, lo} \alpha^t_{vi, v, li, lo} \leq 1 \, &&  \forall v\in V, t\in \mathcal{T} \label{eq:alpha_in_max_1}\\
    &\sum_{vi} \beta^t_{vi, v} \leq 1 &&  \forall v\in V, t\in \mathcal{T} \label{eq:beta_in_max_1}\\
    &\sum_{li, lo} \alpha^t_{vi, vo, li, lo} + \sum_{v, li, lo} \alpha^t_{vo, v, li, lo} && \nonumber \\
    &\leq \sum_{\hat{v}_{2x^\prime - x, 2y^\prime - y}, li, lo} \alpha^t_{vo, \hat{v}, li, lo} + 1 && \forall (vi_{x,y}, vo_{x^\prime,y^\prime}) \in E, t \in \mathcal{T} \label{eq:cornerconflict}
\end{flalign}

The objective function \eqref{eq:obj_func} defines the cost function, which implies that the timestep $t$ is directly proportional to the movement cost.
Therefore, this problem calculates  the flows with the minimum number of steps to convey the target rack from the initial to the goal position.

Constraints \eqref{eq:i_flow_o_flow_alpha}--\eqref{eq:i_flow_o_flow_gamma} are required to satisfy the flow conservation constraints at the vertices.
Racks are conveyed by AGVs, and \eqref{eq:beta_depends_alpha} indicates that the flows of the racks' movements on $G^{+}_{ar}$ are equal to the corresponding flow on $G^{+}_{ag}$.
Constraint \eqref{eq:beta_depends_alpha} indicates that $G^{+}_{ar}$ depends on $G^{+}_{ag}$ and \eqref{eq:gamma_depends_beta} indicates that $G^{+}_{tr}$ depends on $G^{+}_{ar}$.
Considering \eqref{1_gamma_start}, \eqref{eq:i_flow_o_flow_gamma}, and \eqref{eq:gamma_depends_beta}, only flows related to the movement of the target rack on $G^{+}_{ar}$ are reflected in $G^{+}_{tr}$.
Constraint \eqref{eq:alpha_load} indicates that a rack must exist where an AGV loads a rack ($\alpha^t_{v, v, 0, 1}=1$).
Eq. $\sum_{vi} \beta^t_{vi, v}=1$ indicates that one of the flows to vertex $v$ is $1$; that is, a rack is placed at vertex $v$.
Constraints \eqref{eq:alpha_in_max_1} and \eqref{eq:beta_in_max_1} prohibit AGVs and racks from coexisting at the same vertices, respectively.
Constraint \eqref{eq:cornerconflict} is required to prevent the swapping conflict and the corner conflict between agents.

The above formulation is for one target rack. However, the formulation can be extended to multiple target racks.
The network representing the movement of the $i$-th target rack is referred to as $G^{+,i}_{tr}=(V^{+,i}_{tr}, E^{+,i}_{tr})$.
The corresponding flow is $\gamma^{t, i}_{vi, vo}$ and the constraints are the same as $\gamma^{t}_{vi, vo}$.

\subsection{Applications for Real-Time Solving }
The proposed method complicates the network with longer path lengths and the computational cost increases exponentially.
We believe that appropriately dividing the path reduces the computational costs.
Hence, we propose an acceleration method combined with CA*, called \textit{CA*-ILP}. 
This method comprises global and local searches.
The global search finds waypoints to divide the path, and the local search explores the paths between each waypoint.

CA* is an extension of the A* algorithm, tailored for multi-agent scenarios. 
It allows multiple agents to plan their paths sequentially, avoiding collisions through a reservation table that records and manages the paths of all agents. 
This ensures that each agent's path is optimized while preventing overlaps in their paths.

First, path finding is performed using CA* for the global search.
In this step, the racks are assumed to move by themselves, and collisions between racks are allowed.
However, from the viewpoint of timesteps, removing obstacle racks should be avoided.
Therefore, moving to a location with a rack is defined as incurring a $\kappa > 1$ cost in CA*. 
Moving to a location with no racks incurs a cost of $+1$.
Algorithm \ref{algo:ca_global} presents the corresponding pseudo-code.

We define span length $\tau$ between two waypoints before procedure.
A large span length brings the solution closer to the global optimum; however, the computational cost increases.
In the pseudo code, the global search executes CA* [line 1].
It chooses the waypoints according to the global paths [lines 2--7].
The goal locations are also included in the waypoints [line 8].

Second, the local search repeatedly solves the local path-finding problem.
Algorithm \ref{algo:ca_local} presents the pseudo-code.
The local path-finding problem, which involves conveying multiple racks to their waypoints, is solved using ILP [line 3].
The target racks do not always arrive at their waypoint simultaneously.
When one of the racks arrives at its waypoint, a local search is performed again [lines 4--8].

\section{Experiments}
In this section, we describe two experiments: (1) a comparative evaluation of the proposed method against existing methods; and (2) an evaluation of the effectiveness of \textit{CA*-ILP}.
For the experimental setup, the time-expanded networks were represented by NetworkX\footnote{\url{https://networkx.org/}} and the optimization problem was defined by PuLP\footnote{\url{https://coin-or.github.io/pulp/}}.
We used the GUROBI solver\footnote{\url{https://www.gurobi.com/solutions/gurobi-optimizer/}}.
All the experiments were run on a system comprising Ubuntu 22.04, Intel Core i9-12900K, and 128 GiB of RAM.

We performed all the experiments in $6\times4$ grids.
Although this size is too small for automated warehouses, small-sized environments, such as inter-process transportation, are utilized in specific areas of the factories.

\subsection{Experiment 1: Comparative Evaluation of the Proposed Method against Existing Methods}
We compared our initially proposed method, an ILP-based method (independent of CA*), against widely used MAPF solvers, CA* and CBS.
In MARPF, the other racks become obstacles only when the AGV conveys a rack.
We used CA* and CBS for comparison, which accounts for these conditions.
In the first step, the AGVs moved to the initial locations of the target racks without collisions between AGVs.
In the second step, the AGVs moved to the goal locations of the target racks without collisions between AGVs and between racks.

There were initially two AGVs and six racks, including targets. The racks were then added sequentially.
Fig. \ref{exp1_env}(a) illustrates this environment.
Two obstacle racks could block the paths.
For example, two racks were placed in the lower-left region (Fig. \ref{exp1_env}(b)).
We performed 30 experiments at different locations to add the racks.
The solver's time limit was 120 s, and $size_t$ was set to $40$.

\begin{figure}[!t]
  \small
  \begin{algorithm}[H]
    \footnotesize
      \caption{Global Search}
      \label{algo:ca_global}
      \textbf{Input}: Graph $G$, racks ($i$-th rack moves from $n^i_0$ to $n^i_g$)\\
      \textbf{Parameter}: Span length $\tau$ between two waypoints\\
      \textbf{Output}: Sequences of waypoints $waypoints$\\
      \begin{algorithmic}[1]
        \State Find the global paths of the target racks using CA*: $path[i] = (n^i_0, n^i_1, \ldots, n^i_g)$, which is a vertex sequence.
        \ForAll{$i$}
          \State $waypoints[i] = [\,]$
          \State $index = 0$
          \While{$index < |path[i]|-1$}
            \State $waypoints[i] \leftarrow$ add $path[i][current\_index]$
            \State $index = index + \tau$
          \EndWhile
          \State $waypoints[i] \leftarrow$ add $n^i_g$     
        \EndFor 
        \State Return $waypoints$
      \end{algorithmic}
  \end{algorithm}
\end{figure}

\begin{figure}[!t]
  \small
  \begin{algorithm}[H]
    \footnotesize
      \caption{Local Search}
      \label{algo:ca_local}
      \textbf{Input}: Graph $G$, agents, racks, waypoints $waypoints$\\
      \textbf{Output}: Paths of all agents and racks $paths$\\
      \begin{algorithmic}[1]
        \State $paths = [\,]$
        \While{not all racks have arrived at their waypoints}
            \State Use ILP to calculate paths $temp\_paths$ for all target racks from their current position to their respective waypoints
            \State $min\_t$ is the minimum length of $temp\_paths$
            \ForAll{$i$ in all agents and racks}
                \State Update the $paths[i]$ with the $temp\_paths[i][:min\_t+1]$
                \State Update the current position with $temp\_paths[i][min\_t]$
            \EndFor
            \State Check if each rack has reached its waypoint
        \EndWhile
        \State Return $paths$
      \end{algorithmic}
  \end{algorithm}
\end{figure}

Fig. \ref{exp1_suc_rate} shows a comparison of the success rates and average makespans of the successful tasks\footnote{In Fig. \ref{exp1_suc_rate}, there was no difference between CA* and CBS in such a simple environment.}.
As Fig. \ref{exp1_suc_rate} shows, with few added racks, all methods succeeded.
However, with many added racks, CA* and CBS sometimes failed to find the path.
CA* and CBS consider only the movements of the target racks and cannot remove obstacle racks.
Therefore, they could not find the path when many racks were added and the paths were blocked.
Our proposed method considered all rack movements, and as a result, it found the movement necessary for removing obstacle racks.

Although the proposed method is theoretically capable of solving these problems if there are one or more empty vertices, it failed in half of the tasks when adding six racks. 
When the problem was complex and the computational cost was high, the solver failed to find a path within the time limit.
Therefore, we evaluated an acceleration method in the following experiment. 

\begin{figure}[!t]
  \vspace*{1.35mm}
\centering
\includegraphics[scale=0.95]{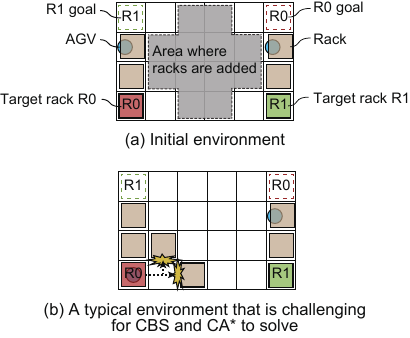}
\caption{Environment of Exp. 1.}
\label{exp1_env}
\end{figure}

\begin{figure}[!t]
\centering
\includegraphics[width=0.75\columnwidth]{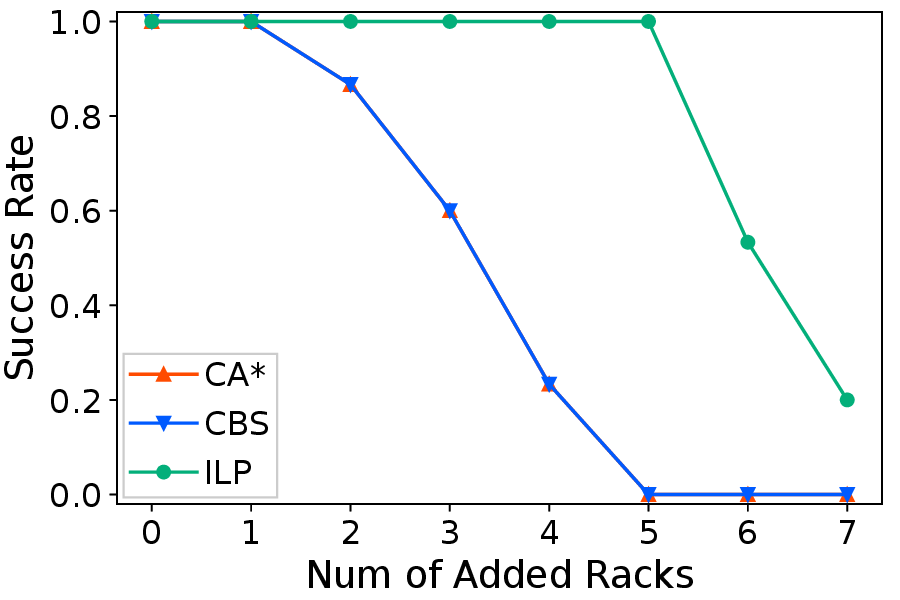}
\caption{Comparison of success rates on Exp. 1.}
\label{exp1_suc_rate}
\end{figure}

\subsection{Experiment 2: Evaluation of the Effectiveness of \textit{CA*-ILP}}
Solving the path-finding problem using the proposed method was computationally expensive.
We confirmed that \textit{CA*-ILP} reduced the computational cost.
We assigned a cost $\kappa=3$ to move to a location occupied by a rack\footnote{We set the cost $\kappa$ to $3$ because we assumed that the task 
involved three steps: loading, moving, and unloading the obstacle rack.}.

\begin{figure}[!t]
  \vspace*{1.35mm}
  \centering
  \includegraphics[scale=0.95]{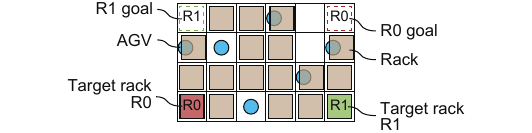}
  \caption{Example of the environment in Exp. 2.}
  \label{exp2_env}
\end{figure}

\begin{figure}[!t]
  \centering
  \includegraphics[width=0.9\columnwidth]{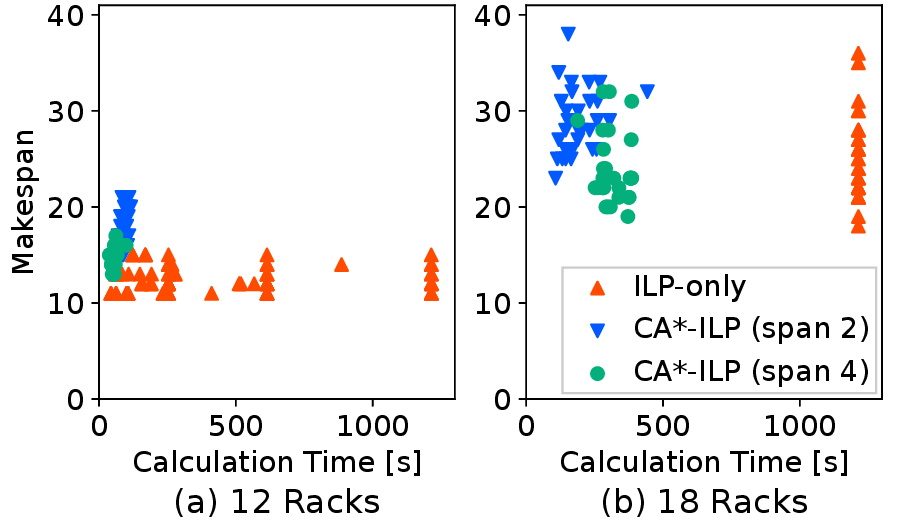}
  \caption{Comparison of makespans on Exp. 2.}
  \label{exp2_makespan}
\end{figure}

The evaluations were conducted in sparse (12 racks) and dense (18 racks) environments.
Each environment contained eight AGVs.
Fig. \ref{exp2_env} is an example of a dense environment.
We performed 30 experiments using the AGVs and non-target racks in different initial locations to compare the makespans of the tasks of conveying the two target racks to their goals.
The comparison of the makespans is shown in Table \ref{exp2_ca_ilp_makespan}, with the differences in span lengths between waypoints (\textit{CA*-ILP}) highlighted.
Fig. \ref{exp2_makespan} also shows a comparison of the makespans between the initially proposed method (\textit{ILP-only}) and the acceleration method (\textit{CA*-ILP}).
We set the time limit of 120s for each local search conducted by \textit{CA*-ILP}, and each $size_t$ was set to $24$.
\textit{ILP-only} was computationally expensive; therefore, we set the time limit to three different values, 120s, 240s, and 1200s, and compared them.
 $size_t$ was set to $40$.
In the dense environment (Fig. \ref{exp2_makespan}(b))\footnote{The calculation time included the time required for network construction.}, \textit{ILP-only} failed to find solutions in some experiments within the time limits of 120s and 240s.

First, we evaluated the difference in span between waypoints (\textit{CA*-ILP}). 
According to Table \ref{exp2_ca_ilp_makespan}, Span 4 yielded the best result.
Span 1, being excessively fine-grained and ignoring future rack locations, had a larger makespan.
We adopted Spans 2 and 4 for the subsequent comparisons.

Second, we compared the initially proposed method (\textit{ILP-only}) with the acceleration method (\textit{CA*-ILP}).
In the sparse environment (Fig. \ref{exp2_makespan}(a)), \textit{CA*-ILP} exhibited a shorter total calculation time compared with \textit{ILP-only}; however, its makespan was larger.
\textit{CA*-ILP} divided the paths into several parts, each easier to solve, thereby reducing the overall calculation cost compared to \textit{ILP-only}.
However, because the waypoints in \textit{CA*-ILP} were not always optimal, its makespan exceeded that of the global optimal solution by \textit{ILP-only}.

\begin{table}[!t]
  \small 
  \centering
  \captionsetup{justification=centering,singlelinecheck=false}
  \caption{Comparison of CA*-ILP makespans on Exp. 2.} 
  \begin{tabular}{l|llll}
    \hline
  Span length $\tau$ between waypoints  & 1         & 2       & 4     \\ \hline
  12 racks   & 21.9     & 18.1    & \underline{14.8}  \\
  20 racks   & 39.9     & 28.9    & \underline{23.9} \\ \hline
  \end{tabular}
  \label{exp2_ca_ilp_makespan}
\end{table}

In the dense environment (Fig. \ref{exp2_makespan}(b)), the makespan of \textit{CA*-ILP} was smaller than \textit{ILP-only}, and the total calculation time of \textit{CA*-ILP} was considerably shorter than \textit{ILP-only}.
In the dense environment, the computation cost was higher than that in the sparse environment, and \textit{CA*-ILP} was more effective in calculating time.
In all the experiments, \textit{ILP-only} found a feasible solution within the time limit of 1200s; however, \textit{ILP-only} did not always find a good feasible solution within the time limit.
Consequently, the average makespan of \textit{CA*-ILP} was smaller than \textit{ILP-only}.

\section{Conclusion}
In this study, we defined the MARPF problem for planning the paths of target racks to their designated locations using AGVs in dense environments without passages.
We proposed an ILP-based formulation for synchronized time-expanded networks by dividing the movements of AGVs and racks.
The proposed method optimized the paths to move the obstacles and convey the target racks.
Based on the complexity that increases with path length, we also presented an acceleration method combined with CA*.
Our experiments confirmed that the acceleration method reduced computational cost.
While we acknowledge the need for further research on more efficient algorithms, it is sufficient if the paths are calculated before the following product is completed  in conveyance between production processes in factories. 
In these instances, our proposed algorithm can be practically applied.

Although we performed our experiments on a small grid, the problem setting of MARPF can be applied to large-scale warehouses.
However, solving bigger problems increases the computational cost.
In the future, we plan to investigate more efficient and faster algorithms.

\section*{Acknowledgments}
We thank Kenji Ito, Tomoki Nishi, Keisuke Otaki, and Yasuhiro Yogo for their helpful discussions.

% Generated by IEEEtran.bst, version: 1.14 (2015/08/26)

\end{document}